\documentstyle[pre,aps,epsfig,twocolumn]{revtex}
\draft 

\def\be{\begin{equation}}
\def\ee{\end{equation}}
\def\bea{\begin{eqnarray}}
\def\eea{\end{eqnarray}}

\begin{document}

\title{Diffusion with random distribution of static traps}

\author{G.~T.~Barkema$^1$, Parthapratim Biswas$^2$, Henk van Beijeren$^1$}
\address{Theoretical Physics$^1$, Utrecht University, Leuvenlaan 4, 3584
CE Utrecht, the Netherlands \\
Debye Institute$^2$, Utrecht University, Princetonplein 5, 3508 TA Utrecht,
the Netherlands}

\date{\today}

\maketitle

\begin{abstract}
The random walk problem is studied in two and three dimensions in
the presence of a random distribution of static traps.  An efficient
Monte Carlo method, based on a mapping onto a polymer model, is
used to measure the survival probability $P(c,t)$ as a function of
the trap concentration $c$ and the time $t$.  Theoretical arguments
are presented, based on earlier work of Donsker and Varadhan and
of Rosenstock, why in two dimensions one expects a data collapse
if $-\ln[P(c,t)]/\ln(t)$ is plotted as a function of $\sqrt{\lambda
t}/\ln(t)$ (with $\lambda=-\ln(1-c)$), whereas in three dimensions
one expects a data collapse if $-t^{-1/3}\ln[P(c,t)]$ is plotted as
a function of $t^{2/3}\lambda$. These arguments are supported by the
Monte Carlo results.  Both data collapses show
a clear crossover from the
early-time Rosenstock behavior to Donsker-Varadhan behavior at long times.
\end{abstract}
\vskip 0.3cm


The problem of trapping in a random medium is fundamental for
understanding many important physical processes in disordered
systems. Practical applications include some dynamical processes in
disordered media, kinetics of reactions, electron-hole recombinations
in random and amorphous solids, exciton trapping and annihilation,
etc.\cite{review}.  In the lattice version of this problem, traps
are randomly distributed on a lattice of dimension $d$.  Usually,
all correlations between traps are ignored, largely motivated by the
fact that in most relevant physical situations the concentration of
traps is small.  Initially at time $t=0$, a large population of random
walkers is uniformly distributed over the lattice. Each walker hops
from one lattice site to a randomly chosen nearest-neighbor site,
at a rate of one hop per time unit. When a walker meets a trap,
it dies.  The time dependence of the survival probability of the
walkers in the presence of traps is an interesting problem, which
has been analyzed both mathematically and numerically in the literature
\cite{review,beeler,kehr1,dv,grass,rs,klafter,fix,havlin,meiro,nieu,luben,anlauf,gallos,holland}.
The problem is related to the properties of the density of states of
the electrons on a lattice with randomly distributed impurities, which
has been discussed in Refs.~\cite{nieu,luben,lifs,lutt}.

Theoretical treatments of the trapping problem usually start by
considering the configuration averaged survival probability $P(c,t)$,
where $c$ is the concentration of static traps. If we assign a label
$i=1\dots N$ to all different random walks of $t$ steps and introduce
$n(i,t)$ as the number of different sites visited by the walk labeled $i$
we can write the survival probability as
\begin{equation}
P(c,t) = \langle (1-c)^{n(i,t)} \rangle,
\label{infT}
\end{equation}
since the probability that a site is not a trap is equal to $(1-c)$.
For short times $P(c,t)$ is well approximated by the Rosenstock (RS)
\cite{rs} expression $(1-c)^{\langle {n(i,t)} \rangle}$. This leads to
exponential dependence on $t$ for $d \ge 3$ and exponential dependence
on $\sqrt{t}$ in $d=1$.  The short-time limit in two dimensions will be
discussed further on in this paper.

The behavior of the survival probability at long times has been treated
rigorously by Donsker and Varadhan \cite{dv}; the limit $t \rightarrow
\infty$ is known in the literature as Donsker-Varadhan (DV) limit. In
the DV limit, the survival probability $P(c,t)$ in $d$ dimensions does
not decay exponentially, as simple intuition would suggest, but in a
rather more complicated way:
\begin{equation}
\ln [P(c,t)] \approx -A \lambda^{\frac{2}{d+2}}\,t^{\frac{d}{d+2}},
\label{pg}
\end{equation}
where $A$ is a constant depending on dimension as well as on the
characteristics of the lattice and the random walk process, and
$\lambda=-\ln(1-c)$, with $c$ the concentration of static traps.
An interpretation of this apparently unusual behavior was given by
Grassberger and Procaccia~\cite{grass}: they assign it to the existence
of very rare large trap-free regions where walkers can survive for a
long time.  With increasing time ever larger trap free regions become
dominant; the probability of finding such regions decreases exponentially
with their $d$-dimensional volume, but the decay rate of particles moving
within such a region is inversely proportional to the square of its
diameter. The optimal choice of this diameter gives rise to the stretched
exponential behavior of Eq.~(\ref{pg}).

One important problem with the DV result is that it does not say anything
about how large $t$ should be, in order to observe this behavior, nor
what should be the corresponding value of the survival probability. The
issue of the onset of the DV behavior has been studied numerically
by a variety of algorithms, resulting in estimates which vary widely,
apparently depending on the kind of algorithms used in the calculation.
Earlier results of Klafter {\em et~al.} \cite{klafter}, based on their
numerical simulations, suggest that $P(c,t)$ should be less than
$10^{-21}$ for $d=2$ to observe this behavior, whereas Fixman \cite{fix}
obtained a value of $P(c,t) < 10^{-67}$ in $d=3$. Among others, Havlin {\em
et~al.} \cite{havlin} used an exact enumeration approach and obtained
$P(c,t) \lesssim 10^{-13}$ for both two and three dimensions.  The scanning
method of Meirovitch \cite{meiro} supported the claims of Havlin {\em
et~al.}  None of the above studies have shown conclusively the crossover
from the early-time RS behavior to the asymptotic DV behavior, except
perhaps at very high trap densities.  The most recent progress comes
from Gallos {\em et~al.} \cite{gallos}, who have studied the problem
numerically in $d=2$, using the idea of self-interacting random walks
coupled with a ``slithering snake'' algorithm. They plotted the quantity
$-t^{-0.1}\ln[P(c,t)]$ as a function of $t^{0.8}\lambda$ and obtained
a fairly satisfactory collapse, within their numerical precision; the
theoretical basis for this kind of data collapse is however 
unclear.

Our algorithm for the symmetric random walk on a simple cubic or
quadratic lattice, the case studied mostly before, is also based on
the identification of each random walk of $t$ steps with a polymer
configuration of length $t$.  A direct Monte Carlo procedure to
estimate the survival probability would then be to generate a large
number of random walks, and average the quantity $(1-c)^{n(i,t)}$. This
procedure would however give poor statistics with increasing $t$, since
the result would be dominated by the one or two random walks that are
the most compact. More accurate Monte Carlo results can be obtained by
biasing the random walks generated towards more compact ones. Here, the
identification of random walks with polymers comes into play. To each
polymer configuration $i$ we attribute an energy, given by the Hamiltonian
\begin{equation}
H(i,t)= \lambda n(i,t).
\label{hamil}
\end{equation}
If we then sample polymer configurations at infinite temperature, the
average energy is equal to $\lambda n(i,t)$, which in turn equals
$\ln[P(c,t)]$. Numerically more accurate results are obtained, however,
by generating polymer configurations at some inverse temperature
$\beta$.  The likelihood for a configuration $i$ to be generated is
then proportional to its Boltzmann factor $\exp(-\beta H(i,t))$ and as
a consequence there is a bias towards compact configurations. Also from
this ensemble, one can compute the survival probability using
\begin{equation}
P(c,t) = \frac{\langle \exp(+\beta H(i,t)) (1-c)^{n(i,t)} \rangle_{\beta}}
                 {\langle \exp(+\beta H(i,t)) \rangle_{\beta}},
\end{equation}
where the averaging is performed over the ensemble generated at inverse
temperature $\beta$. Note that at infinite temperature, i.e. $\beta=0$,
there is no bias and we retrieve Eq.(\ref{infT}). Interestingly, at inverse
temperature $\beta=1$ the numerator equals unity and the inverse of the
survival probability equals the average of the inverse Boltzmann weights.
It turns out that the most accurate results are obtained at some inverse
temperature between 0 and 1; all our results are obtained at $\beta=0.75$.

\begin{figure}
\epsfxsize=3.0in
\epsfbox{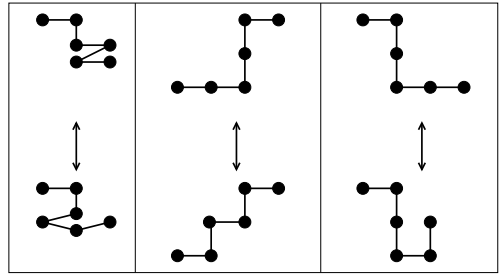}
\caption{Monte Carlo moves used in the simulation: exchange of a pair
of consecutive steps (left and middle), or replacement of the first or
last step (right).}
\label{fig:MCmoves}
\end{figure}

We generate a Markov chain of polymer configurations at inverse temperature
$\beta$ by proposing randomly chosen single-monomer moves, and accepting
each proposed move with the Metropolis acceptance probability
\begin{equation}
P_{acc}=\min \left[ 1,\exp(-\beta \Delta E) \right], 
\end{equation}
where $\Delta E$ is the change in energy, defined by the Hamiltonian
Eq. (\ref{hamil}). The types of single-monomer moves that we propose
are depicted in Fig.~\ref{fig:MCmoves}. In a naive implementation,
the computational effort per move would scale linearly with polymer
length because of the necessity to walk along the polymer to compute
the change in energy. However, if one stores in each lattice site
the number of times the polymer visits that site, then the energy
difference can be computed easily: the energy increases by $\lambda$
if the monomer moves to a site which is not visited before, decreases by
$\lambda$ if the monomer leaves a site that it visits only once, and
stays unaltered if either both or neither of these two situations hold.
This administration is also easily updated: the new site is visited one
more time, while the old site is visited one time less.

\begin{figure}
\epsfxsize=3.0in
\epsfbox{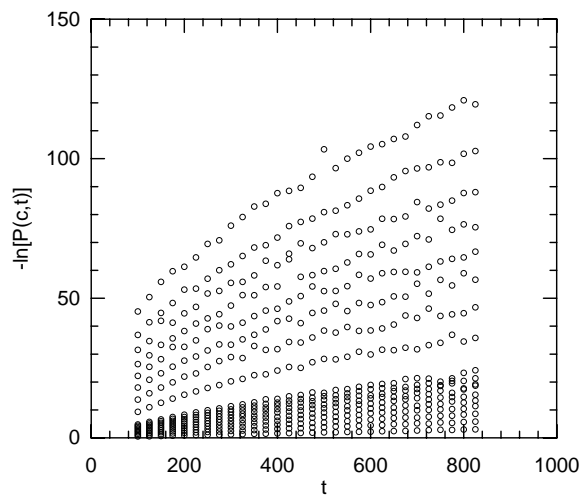}
\caption{Monte Carlo results for the survival probability as a function
of time, for several trap concentrations on a two-dimensional lattice:
$-\ln[P(c,t)]$ is plotted as a function of time $t$.}
\label{fig:raw}
\end{figure}

We now discuss numerical results obtained using our algorithm outlined
above. First, we concentrate on the case of two dimensions. In
Fig.~\ref{fig:raw}, we have plotted the logarithm of the survival
probability $P(c,t)$ for concentrations $c=0.01, 0.02, \dots, 0.09;
0.1, 0.2, \dots, 0.9$, and times (polymer lengths) up to $t=825$. In
these simulations, we start with a random walk, which is then simulated
over $10^9$ Monte Carlo steps per monomer, at an inverse temperature of
$\beta=0.75$. Data obtained during the first $50000t$ attempted Monte
Carlo steps per monomer have been discarded (thermalization).

We now obtain a theoretically justified data collapse based on the
identification of a common scaling variable for the DV and RS regimes.
At short times and small concentrations, we expect that the polymer
configuration closely resembles a random walk. In the case of two
dimensions, this means that the number of different sites visited scales
as $t/\ln(t)$~\cite{montroll}.  Consequently, we deduce that the RS behavior
in two dimensions is properly described by
\begin{equation}
-\ln[P(c,t)] \sim \lambda t/\ln(t).
\label{RS2}
\end{equation}
At long times and large concentrations we expect to observe the DV behavior:
\begin{equation}
-\ln[P(c,t)] \sim \left( \lambda t\right)^{1/2}.
\label{DV2}
\end{equation}
To obtain the proper scaling variable, we equate the right terms in equations 
(\ref{RS2}) and (\ref{DV2}), and obtain $\sqrt{\lambda t}/\ln(t)=1$; we can use 
this as a scaling variable, and rewrite equations (\ref{RS2}) and (\ref{DV2}) 
as:
\begin{eqnarray}
\frac{-\ln[P(c,t)]}{\ln(t)} &=& \left[ \frac{\sqrt{\lambda t}}{\ln(t)} 
\right]^2;\nonumber\\
\frac{-\ln[P(c,t)]}{\ln(t)} &=& \frac{\sqrt{\lambda t}}{\ln(t)}.
\end{eqnarray}

Thus we expect that if $-\ln[P(c,t)]/\ln(t)$ is plotted as a function of
$\sqrt{\lambda t}/\ln(t)$, the data for all values of trap concentration
$c$ and time $t$ collapse onto a single curve, with an effective exponent
that crosses over from 2 to 1.

\begin{figure}
\epsfxsize=3.0in
\epsfbox{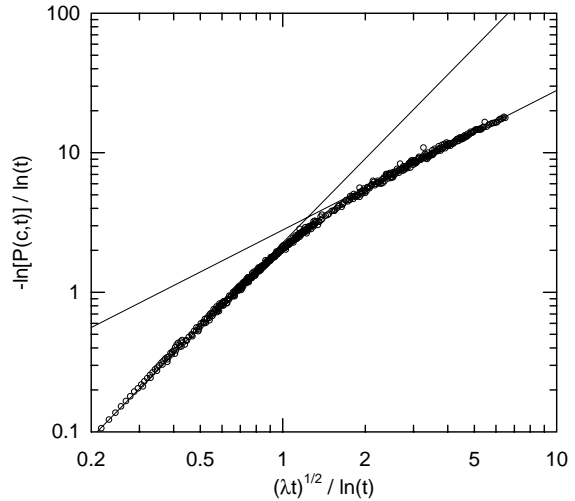}
\caption{Collapse of the two-dimensional data: $-\ln[P(c,t)]/\ln(t)$ is
plotted as a function of $\sqrt{\lambda t}/\ln(t)$ in a double-logarithmic
plot. The solid lines are fits to the data, with slopes 2 and 1. They
cross at the point (1.13, 3.5).}
\label{fig:collapse2D}
\end{figure}

In Fig.~\ref{fig:collapse2D} we have performed this plot, using the same
data as in Fig.~\ref{fig:raw}.  Clearly, the data collapse is convincing
over the whole range of parameters used in our simulation.  Also, one can
clearly identify the RS regime with a slope of 2 and the DV regime with
a slope of 1, both indicated with solid lines.  The numerical estimate
of the point where these lines cross is $-\ln[P(c,t)]/\ln(t)=3.5$ and
$\sqrt{\lambda t}/\ln(t)=1.13$.  The survival probability where the DV
regime starts is then given by $-\ln[P(c,t)]=3.1 \sqrt{\lambda t}$, for
a suitable choice of $c$ and $t$.  The fact that the survival probability
at the start of the DV regime is not a constant explains the wide range of
reported values for this quantity.  The apparent dependence of the results
on the simulation methods can be understood because certain simulation
methods are especially suited for high $c$-values whereas other methods
require fairly low values of $c$. From our results one may conclude that
the DV regime can already start at a significant survival probability,
depending on $c$; it should therefore be observable experimentally.

\begin{figure}
\epsfxsize=3.0in
\epsfbox{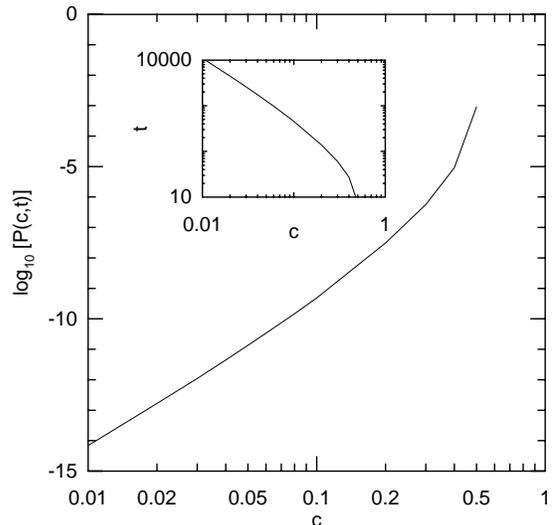}
\caption{Location of the crossover probability $P(c,t)$ as a function of
trap concentration $c$, in two dimensions.  The inset shows the crossover
time $t$ as a function of $c$.}
\label{fig:cross2d}
\end{figure}

To illustrate this in detail we plotted in Fig.~\ref{fig:cross2d} the cross-over 
values of both time and survival probability as functions of the trap 
concentration.

In dimensions different from
$d=2$ it is equally possible to define a common scaling variable for the
RS and DV regimes, that ought to give rise to a data collapse. For
$d<2$ one should consider $\ln[P(c,t)]$ as a function of $(\lambda t)^{d/2}$
and for $d>2$ collapse should occur for $-t^{(2/d-1)}\ln[P(c,t)]$ as
a function of $t^{2/d}\lambda$. From these expressions one first of
all sees very clearly that $d=2$ acts as a cross-over value for the
dimensionality, as suggested already by the logarithmic terms in the
$2d$ scaling variables.

In particular, in three dimensions, where the number of different sites
visited increases linearly with time $t$, the RS and DV regimes can be
written as:

\begin{eqnarray}
-t^{-1/3}\ln[P(c,t)] &=& t^{2/3}\lambda;\nonumber\\
-t^{-1/3}\ln[P(c,t)] &=& \left[ t^{2/3}\lambda \right]^{2/5}.
\end{eqnarray}

\begin{figure}
\epsfxsize=3.0in
\epsfbox{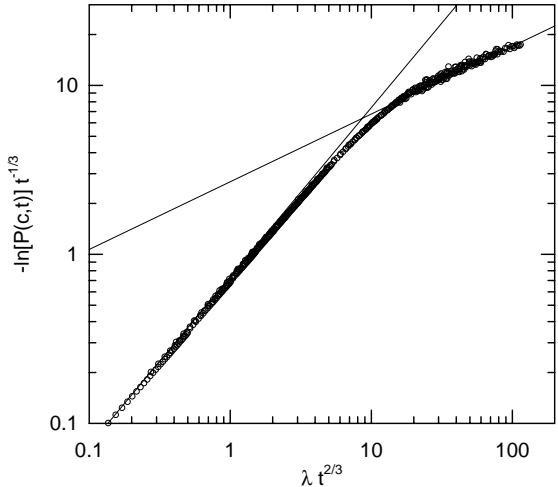}
\caption{Collapse of the three-dimensional data: $-t^{-1/3}\ln[P(c,t)]$ is
plotted as a function of $t^{2/3}\lambda$ in a double-logarithmic
plot. The solid lines are fits to the data, with slopes 1 and 2/5. They
cross at the point (8.5, 6.3).}
\label{fig:collapse3D}
\end{figure}

To show the resulting data collapse in three dimensions, we plotted
in Fig.~\ref{fig:collapse3D} $-t^{-1/3}\ln[P(c,t)]$ as a function of
$t^{2/3}\lambda$.  All data have been obtained again with the polymer
algorithm described above.  As in two dimensions, they can be explained
by a crossover from the RS regime, where the curve has a slope of 1,
to the DV regime, where the curve has a slope of 2/5.

\begin{figure}
\epsfxsize=3.0in
\epsfbox{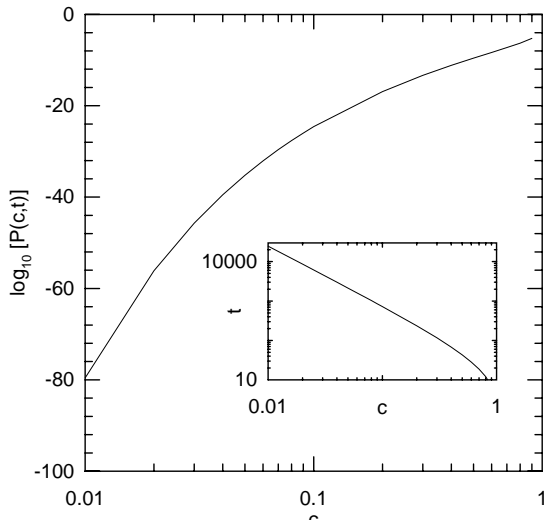}
\caption{Location of the crossover probability $P(c,t)$ as a function of
trap concentration $c$, in three dimensions.  The inset shows the crossover
time $t$ as a function of $c$.}
\label{fig:cross3d}
\end{figure}

In Fig.~\ref{fig:cross3d} we also plotted the cross-over time and survival
probability as functions of $c$ for three dimensions. As one might expect, the 
concentration dependence is even much stronger than in two dimensions.

In summary, we have outlined an efficient algorithm to investigate
diffusion with random traps in an arbitrary concentration range.
The simulation results for all trap concentrations and times could
be explained by a crossover from the Rosenstock behavior at short times
to the Donsker-Varadhan behavior at long times. We have identified
the location of the crossover point, which in contrast to earlier belief,
does not simply yield a specific value for the survival probability.

\end{document}